\documentclass[12pt]{article}

\usepackage{scicite}

\usepackage{times}

\usepackage{graphicx,color}
\usepackage{epstopdf}
\usepackage{verbatim}
\usepackage{siunitx}
\usepackage{amsmath}
\usepackage{gensymb}

\usepackage{ulem}

\newenvironment{sciabstract}{%
\begin{quote} \bf}
{\end{quote}}

\title{Spontaneous oscillations and negative-conductance transitions in microfluidic networks} 

\author
{Daniel J. Case,$^{1}$ Jean-R\'egis Angilella,$^{2}$ and Adilson E. Motter$^{1,3\ast}$\\
\\
\normalsize{$^{1}$Department of Physics and Astronomy, Northwestern University, Evanston, IL 60208, USA}\\
\\
\normalsize{$^{2}$Normandie Universit\'e, UNICAEN, UNIROUEN, ABTE, Caen 14000, France}\\
\\
\normalsize{$^{3}$Northwestern Institute on Complex Systems, Northwestern University, Evanston, IL 60208, USA}\\
\\
\normalsize{$^\ast$To whom correspondence should be addressed; E-mail:  motter@northwestern.edu.}\\
\\
}

\date{}

\begin{document}

\baselineskip24pt

\maketitle 

\noindent
\normalsize{NOTICE: This is the authors' version of the work. It is posted here by permission of the AAAS for personal use, not for redistribution. The definitive version was published in Science Advances 6, eaay6761 (2020), DOI: 10.1126/sciadv.aay6761.}\\

\begin{sciabstract}
 The tendency for flows in microfluidic systems to behave linearly poses a challenge for designing integrated flow control 
schemes to carry out complex fluid processing tasks. This hindrance has led to the use of numerous external control devices to manipulate flows, thereby thwarting the potential scalability and portability of lab-on-a-chip technology. Here, we devise a microfluidic network exhibiting nonlinear flow dynamics that enable new mechanisms for on-chip flow control. 
This network is shown to exhibit oscillatory output patterns, bistable flow states, hysteresis, signal amplification, and negative-conductance transitions, 
all without reliance on dedicated external  control hardware, movable parts, flexible components, or oscillatory inputs. These dynamics arise from nonlinear fluid inertia effects in laminar flows that we amplify and harness through the design of the network geometry. 
We suggest that these results, which are supported by fluid dynamical simulations and theoretical modeling, have the potential to inspire development of new built-in control  capabilities, such as on-chip  timing and synchronized flow patterns. 
\end{sciabstract}

\section*{Introduction}
Microfluidic systems---networks of miniature flow channels capable of processing fluids---are now commonly used 
in applications ranging from chemical analysis \cite{Geertz2012a} and flow cytometry \cite{McKenna2011} to computing \cite{Katsikis2015} and point-of-care diagnostics \cite{Sackmann2014}. 
The value of microfluidic networks is manifest in their utility for manipulating fluid motion with precision.  
However, such manipulation is often controlled through the use of external hardware \cite{Pennathur2008,Sackmann2014,Volpatti2014}.
For instance, microscopic valves generally need to be actuated by macroscopic, computer-operated pumps \cite{Unger2000}, 
which has impeded development of portable microfluidic systems \cite{Sackmann2014,Volpatti2014}.
The need for active control stems from the low Reynolds numbers typical of microfluidic flows, whereby fluid inertia forces are small relative to viscous dissipation, causing flow rate changes to be linearly related to pressure changes \cite{Squires2005}.
Thus, it remains challenging to design integrated control mechanisms that are capable of inducing responsive flow dynamics, such as oscillations, switching, and amplification, without relying on  nonlinear input signals or moveable parts.

Nonetheless, significant progress has been made in the development of built-in microfluidic controls.
State-of-the-art approaches for incorporating passive valves for flow-rate regulation generally take advantage of flexible membranes and surfaces to generate nonlinear fluid-structure interactions \cite{Seker2009,Weaver2010,Mosadegh2010}.  
Complex flow patterns and operations have been implemented in such networks, but flexible components can hinder integration, yield to high driving pressures, and may require polymer materials that are not chemically compatible with the working fluid \cite{Whitesides2006,Volpatti2014, Sackmann2014}.
On the other hand, recent appreciation has emerged for the impact and utility of fluid inertia effects on manipulating local flow dynamics in microfluidics \cite{Zhang2016,Stoecklein2019}.
It has been shown that even for moderate Reynolds numbers, the formation of vortices and secondary flows can be 
exploited for particle segregation \cite{DiCarlo2009,Chen2015,Wang2015,Rafeie2016}, mixing fluids \cite{Sudarsan2006, Haward2016}, and diverting flow streams \cite{Tesar2011,Amini2013}.

Here, we present a microfluidic network construction that demonstrates new dynamics resulting from fluid inertia, which, importantly, can serve as novel flow control mechanisms and facilitate the design of  integrated microfluidic systems.
Our network exhibits: 
(i) spontaneous emergence of persistent flow-rate oscillations for fixed driving pressures;
(ii) hysteretic flow behavior in which more than one set of stable flow rates exist for the same driving pressures; and
(iii)  negative-conductance \textit{transitions}, whereby an increase (decrease) in the driving pressure leads to a discontinuous decrease (increase) in the flow rate. 
These behaviors are interesting in their own right, and are analogous to 
 to behaviors formerly sought through different approaches.
Oscillations have been implemented in microfluidic networks by using flexible components \cite{Lammerink1995,Tas2002} and  utilized as a timing mechanism \cite{Duncan2013, Mosadegh2010}.
Hysteresis has been explored through the implementation of hysteretic valves and, along with oscillatory driving, found applications in establishing microfluidic logic systems \cite{Rhee2009,Weaver2010}. 
 Non-monotonic pressure-flow relationships analogous to negative-conductance transitions,  have been previously sought using flexible diaphragm valves \cite{Xia2017,Gomez2017} and used for signal amplification and flow switching.
Our network does not include flexible components nor relies on   oscillatory inputs. Instead, the behaviors in (i)-(iii) arise 
by structuring the network so that dynamic vortices are generated in the flow and
 nonlinear fluid inertia effects are amplified.
The results presented in this work are derived from simulations of the Navier-Stokes equations and an analytical dynamical model developed to capture the diverse flow properties of the network.

\begin{figure*}[htb] 
\includegraphics[width=0.99\textwidth]{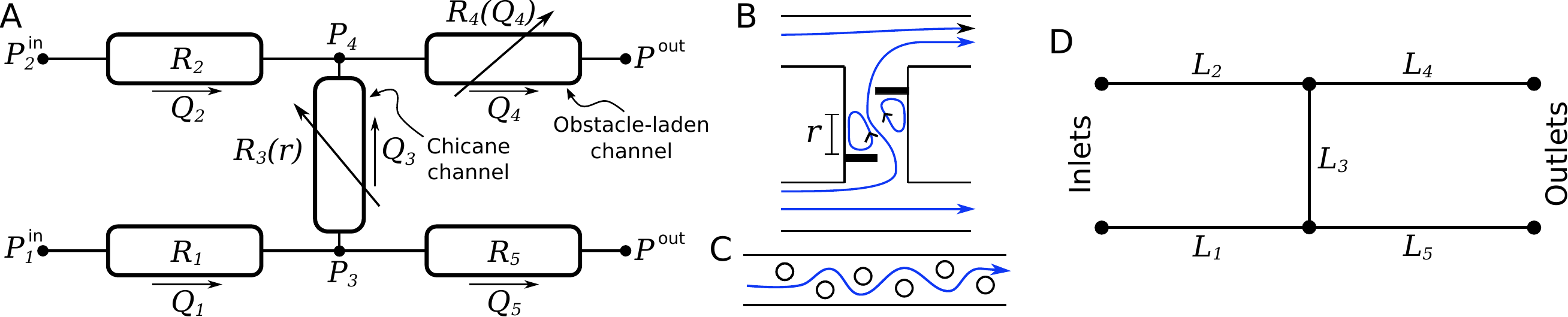}
\caption{
\textbf{Microfluidic network structure.} (\textbf A) Circuit schematic of the network, where the labels denote pressures ($P_i$), channel resistances ($R_i$), and flow rates ($Q_i$).
The inlet and outlet pressures are identified by the superscripts ``in" and ``out", respectively, and the positive flow directions are indicated by arrows.
Two channels exhibit variable (flow-dependent) resistance due to the presence of obstacles. (\textbf{B-C}) Geometric structure of the chicane  (B) and obstacle-laden (C) channels. The blue curves mark example streamlines and specify flow direction. The closed streamlines in (B) represent vortices that form near the barriers and $r$ marks the linear size of the left vortex.
(\textbf D) Network topology of the circuit in (A), where the length of each channel segment is labeled by $L_i$.
}
\label{fig1}
\end{figure*}

\section*{Microfluidic network description and simulation results}
\vspace{-1.75mm}
A circuit schematic of our microfluidic network is shown in Fig.~\ref{fig1}.
The network consists of five channel segments that are constructed into two parallel paths connected by a transversal path. 
Generally, the steady-state relation between the flow rate $Q$ through a microfluidic channel and the pressure loss $\Delta P$ along the channel  takes the form $\Delta P = R Q$, where $R$ is the (absolute) fluidic resistance of the channel. When $R$ is constant, this relation is analogous to Ohm's law for electronic resistors \cite{Oh2012,Perdigones2014}. 
Therefore, we represent the three 
(straight) channels in the network that exhibit constant fluidic resistance
as linear resistors in the schematic (Fig.~\ref{fig1}A).
The two remaining channels include either a chicane of blade-like barriers (Fig.~\ref{fig1}B) or an array of six cylindrical obstacles (Fig.~\ref{fig1}C) that induce nonlinear pressure-flow relations and are represented as nonlinear resistors. 
As we show below, the obstacle-laden channel serves to amplify inertial effects 
and the chicane channel gives rise to oscillations. 
The lengths of the channels vary (Fig.~\ref{fig1}D) but all share a common width $w$ of $500\,\mu$m.
The cylindrical obstacles have a radius of $w/5$ and the two barriers, which extend to the center of the chicane channel, are of thickness $w/10$. No-slip boundary conditions are assumed at all surfaces,
and we consider the static pressure at the outlets of the system $P^{\mathrm{out}}$ to be held at a fixed common value, taken to be zero. At the inlets, we control either the pressures  ($P^{\mathrm{in}}_1$ and $P^{\mathrm{in}}_2$) or the flow rates ($Q_1$ and $Q_2$).

We present the outstanding properties of this microfluidic network through fluid dynamics simulations of incompressible flow in two-dimensions. 
We consider a water-like working fluid with density $\rho=1000$ kg/m$^3$ and dynamic viscosity $\mu = 10^{-3}$~Pa$\cdot$s.
In microfluidics, pressure-driven flow is used across a variety of applications \cite{Pennathur2008}, whereby the system inlets are connected to a pressurized fluid reservoir, the outlets are open to atmosphere (or a lower pressure reservoir), and flow is driven by the resulting pressure gradient. 
Here, we investigate the case in which a common static pressure is applied at the inlets, that is $P^{\mathrm{in}}_1 =P^{\mathrm{in}}_2 = P^{\mathrm{in}}$, which corresponds to the physical scenario in which the inlets are connected to a high pressure reservoir through intermediate passive pressure regulators. 

\begin{figure}[htb] 
\centering
\includegraphics[width=0.45\columnwidth]{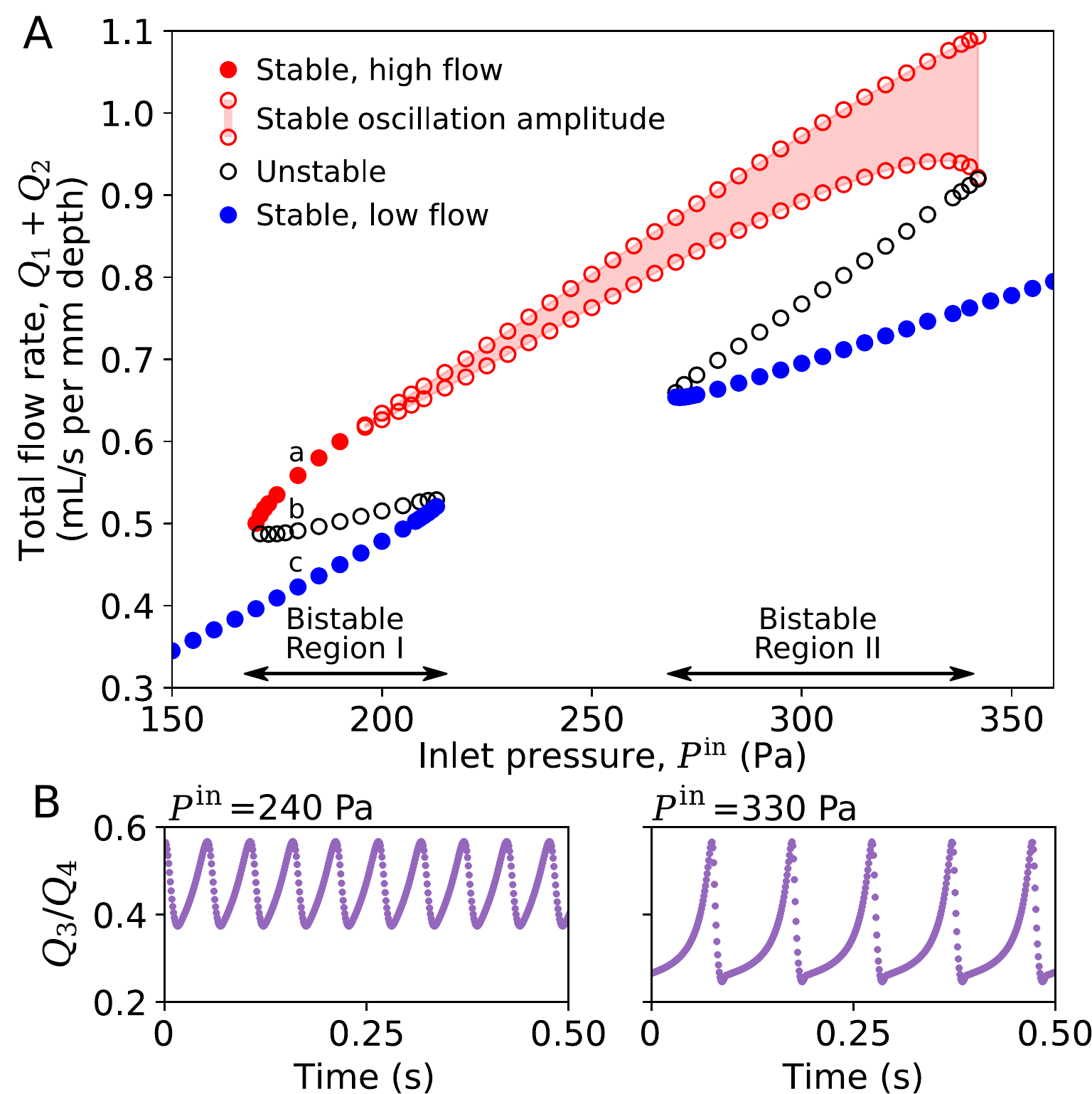}
\caption{
\textbf{Bistability and spontaneous oscillations.} (\textbf{A}) Bifurcation diagram of total flow rate as a function of the inlet pressure $P^{\mathrm{in}}$, generated from direct simulations of the network in Fig.~\ref{fig1} for $P^{\mathrm{in}}_1 =P^{\mathrm{in}}_2 = P^{\mathrm{in}}$.
There exist stable high-flow (red) and low-flow (blue) solution branches, separated by an unstable intermediate-flow branch (black). Oscillating solutions arise spontaneously along the high-flow branch, where the oscillation amplitude is indicated by the shaded region.
The Reynolds number for flows through the chicane channel and the obstacle-laden channel are in the range of $14$--$90$ and $80$--$155$, respectively.
The solutions for $P^{\mathrm{in}}=180\,$Pa, marked with a, b, and c, will be used as references in comparing with other figures. 
The unstable solutions are determined through flow-controlled simulations (see supplementary materials and fig.~S1).
(\textbf{B})
Time series of the proportion of flow exiting the obstacle-laden channel that passes through the chicane channel ($Q_3/Q_4$) for two driving pressures that yield oscillatory flows, showing that frequency decreases and amplitude increases as the driving pressure is increased.
\vspace{-1mm}
}
\label{fig2}
\end{figure}

In Fig.~\ref{fig2}A, we show simulation results of the total flow rate $Q_T = Q_1 + Q_2$ through the network in Fig.~\ref{fig1} over a range of  driving pressures, $P^{\mathrm{in}}$, from which we observe two striking properties.
First, for $P^{\mathrm{in}}$ within two disjoint ranges, two stable solutions for the total flow rate exist.
Second, a subset of solutions are unsteady and exhibit oscillating flow rates (supplementary materials, fig.~S3, and Movie S1),
despite $P^{\mathrm{in}}$ being fixed.
In particular, we find that at a critical value of $P^{\mathrm{in}}$, solutions along the high-flow branch (red symbols in Fig.~\ref{fig2}A) become small-amplitude limit cycles. The corresponding amplitudes and periods grow with $P^{\mathrm{in}}$ (the frequency of the oscillations range from 4 to 20 Hz; see supplementary materials and fig.~S2), and at a higher critical $P^{\mathrm{in}}$ the limit cycle collides with the unstable branch, thereby destabilizing the high-flow solution branch.
An important property of the oscillating solutions is that the proportions of the flow rates through different channel segments also become time-dependent (Fig.~\ref{fig2}B).
Bistability and spontaneous oscillations have been previously studied in 
fixed-structure microfluidic networks 
when feedback loops are incorporated \cite{Khelfaoui2009} or when multiple working fluids with different viscosities are used \cite{Storey2015}.
However, neither of these mechanisms are required in our system.

\begin{figure}[htb]
\centering 
\includegraphics[width=0.65\columnwidth]{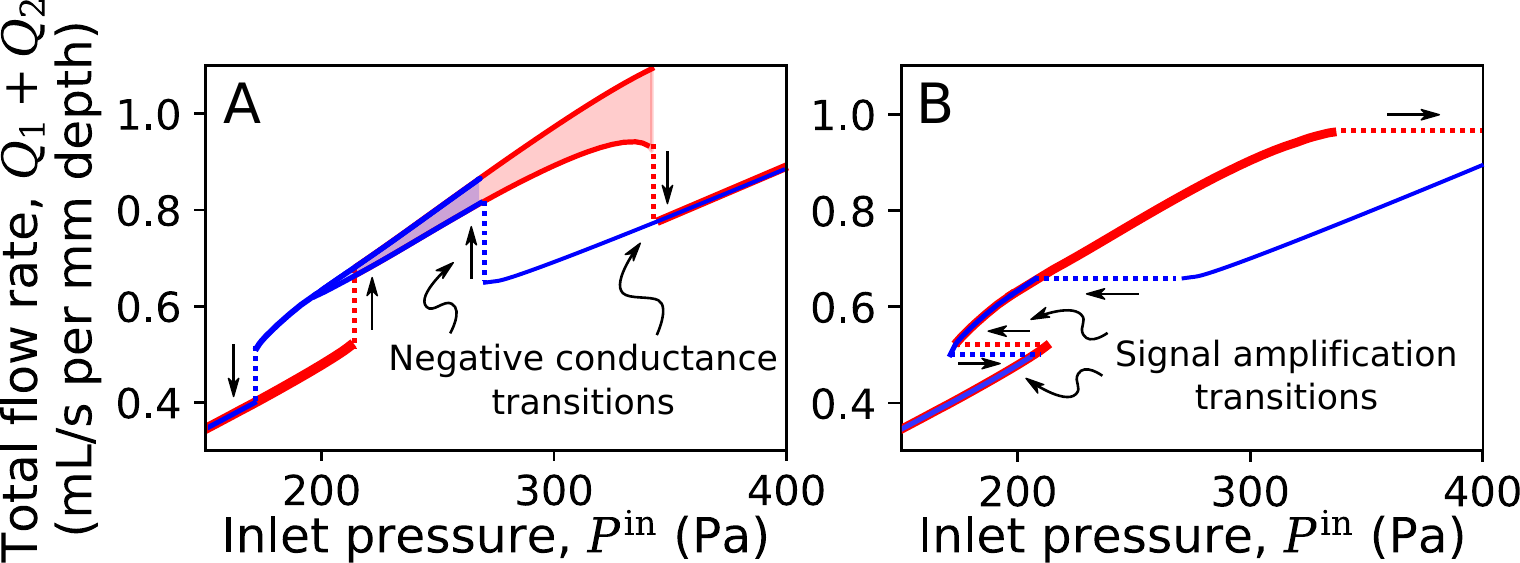}
\caption{
\textbf{Hysteresis and flow state transitions.}
(\textbf{A}) Hysteresis loop and resulting negative-conductance transitions for the network in Fig.~\ref{fig1} when quasistatically increasing (red) or decreasing (blue) the inlet driving pressure. (\textbf{B}) Counterpart of  (A) and resulting signal amplification transitions when quasistatically varying the total flow rate. 
For the latter, $Q_1$ and $Q_2$ are controlled so as to maintain equal pressures at the inlets.
}
\label{fig3}
\end{figure}

Another outstanding property that arises from the bistability in our system is the possibility of negative-conductance transitions and other sudden transitions in $Q_T$ that result from small changes in $P^{\mathrm{in}}$. We characterize these transitions, which occur at the boundaries of the bistable regions (Fig.~\ref{fig2}A), by defining (local) fluidic conductance and resistance as $C=\delta Q_T/\delta P^{\mathrm{in}}$ and its reciprocal, respectively. Here, $\delta$ indicates a finite change and $P^{\mathrm{in}}$ is the controlled variable. Therefore, negative-conductance and negative-resistance  transitions occur when an increase (decrease) in $P^{\mathrm{in}}$ leads to a decrease (increase) in $Q_T$. More importantly, our system exhibits transition points, as shown in Fig.~\ref{fig3}A, at which $C(\delta P^{\mathrm{in}})$ diverges in the limit of small $\delta P^{\mathrm{in}}$: two points at which $C(\delta P^{\mathrm{in}} \rightarrow 0)=+\infty$, corresponding to positive-conductance transitions, {\it and} two points at which $C(\delta P^{\mathrm{in}} \rightarrow 0)=-\infty$, corresponding to negative-conductance transitions. Figure~\ref{fig3}{B} shows that related transitions emerge when the flow rate  $Q_T$, rather than the pressure $P^{\mathrm{in}}$, is taken as the control variable. In this case, a change in $Q_T$ can lead to transitions in which $P^{\mathrm{in}}$ changes by a finite amount. In particular, the later includes signal amplification transitions, which are remarkable transitions in which an infinitesimal increase (decrease) in $Q_T$ leads to a finite decrease (increase) in $P^{\mathrm{in}}$. Both the pressure and flow driven transitions reported here are intimately related to the emergence of hysteresis in the system, which is another consequence of bistability that has potential applications in the development of systems with built-in memory.

\begin{figure}[htb] 
\centering
\includegraphics[width=0.5\columnwidth]{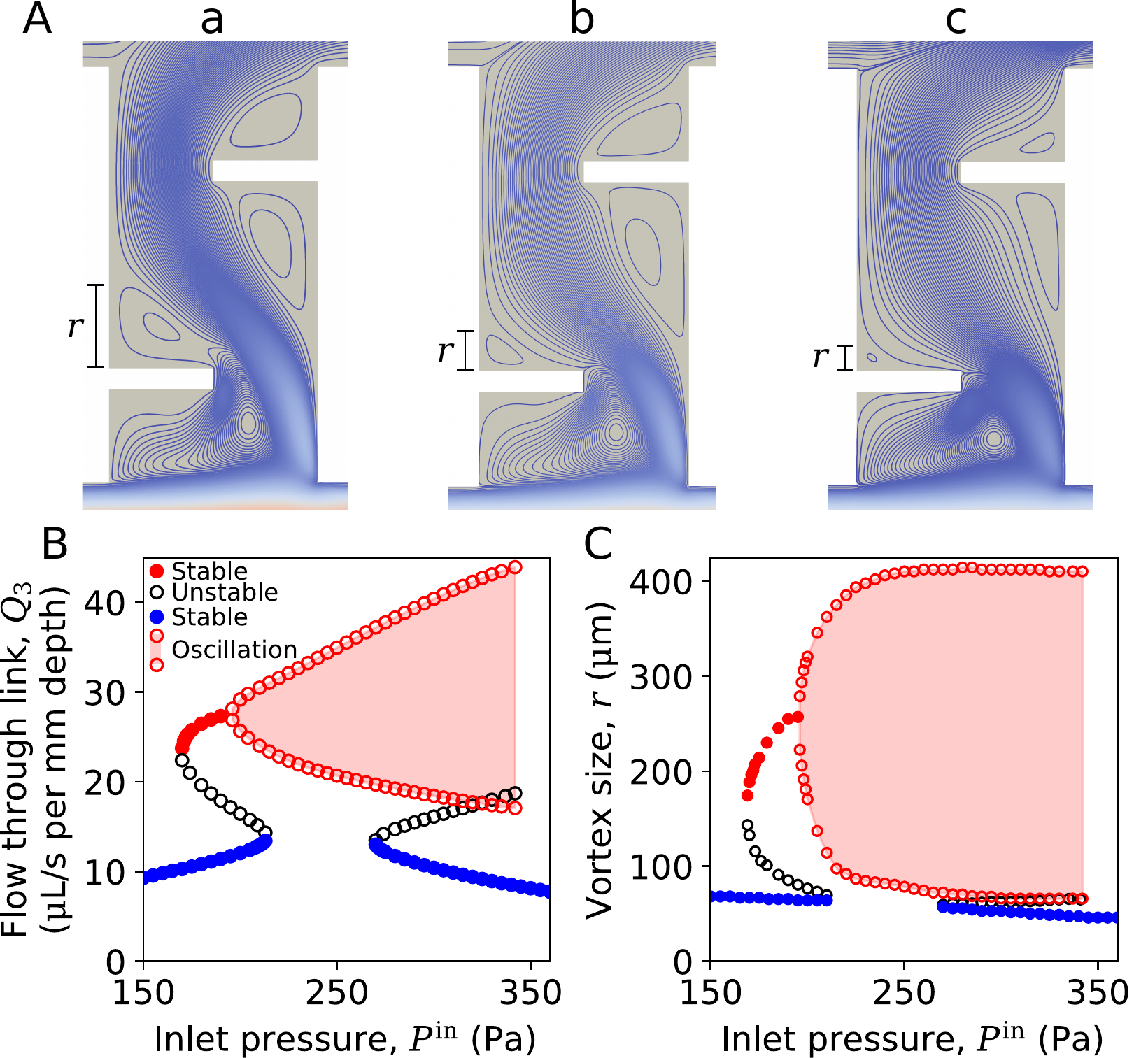}
\caption{
\textbf{Flow structure in chicane channel.}
(\textbf A) Streamlines corresponding to the labeled solutions in Fig.~\ref{fig2}A show variations in the vortices around the blade barriers. 
The size of one of the vortices is denoted by $r$.
(\textbf{B-C}) Bifurcation diagrams for $Q_3$ (B) and $r$ (C), corresponding to all simulation results presented in Fig.~\ref{fig2}A.}
\label{fig4}
\end{figure}

The solutions belonging to the different branches in Fig.~\ref{fig2}A can be further distinguished by the flow rates through specific channels as well as the internal flow structure. It is particularly insightful to examine the streamlines around the complex geometry in the chicane channel, and the associated flow rate $Q_3$. In Fig.~\ref{fig4}A, we show the streamlines corresponding to the three labeled states in Fig.~\ref{fig2}A. A number of steady vortices are observed in the flow around the barriers. The sizes of the vortices are correlated and we designate $r$ to be the size of one of them, as labeled in Figs.~\ref{fig1}B and \ref{fig4}A. We use a one-dimensional measure for $r$, taken to be the distance from the barrier to the vortex reattachment point along the channel wall. 
In  Fig.~\ref{fig4}, B and C, we show that both $Q_3$ and $r$ differ markedly for solutions belonging to the three  branches in Fig.~\ref{fig2}A and that oscillations simultaneously emerge in these variables (supplementary materials, fig.~S3, and Movie S1). Notably, solutions along the high- (low) flow branch in Fig.~\ref{fig2}A correspond to large (small) values of $Q_3$ and $r$.

\begin{figure*}[htb] 
\centering
\includegraphics[width=0.99\textwidth]{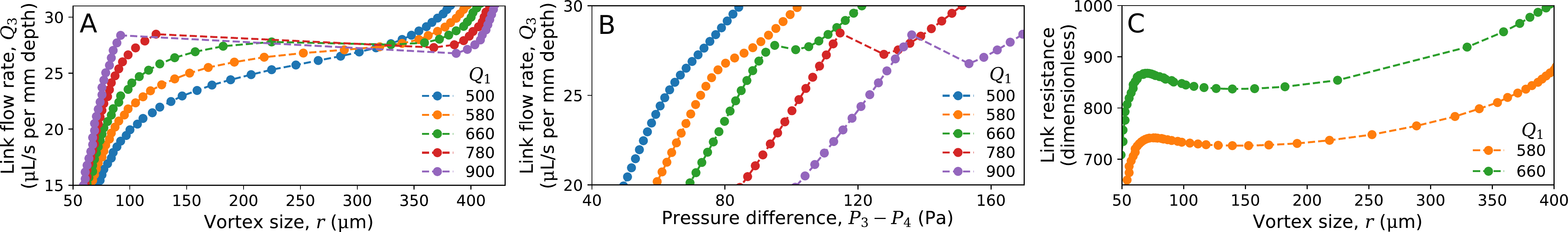}
\caption{
\textbf{Vortex-flow rate interaction.}
(\textbf{A-C}) Navier-Stokes simulation results of the network in Fig.~\ref{fig1} for fixed values of $Q_1$ as $Q_2$ is increased, from which we determine the relation between $Q_3$ and $r$ (A), the pressure-flow relation for the chicane channel (B), and the dependence of the chicane channel resistance on $r$ (C). 
Transitions are evident at the points of discontinuity in (B), which can be associated with the points of discontinuity in Fig.~\ref{fig3}, albeit for different control and independent variables.  
The pressures $P_3$ and $P_4$ are approximated from simulations by averaging pressure values sampled across the channel width near the chicane channel junctions.
The chicane channel resistance is defined as ($P_3-P_4)/Q_3$ and is non-dimensionalized by dividing it by $\mu/w^2$.
}
\label{fig5}
\end{figure*}

We determine the relationship between  $r$ and $Q_3$ by performing simulations in which the flow rates at both inlets ($Q_1$ and $Q_2$) are controlled.
From these simulations we compute $r$, $Q_3$, and the pressure loss along the chicane channel $\Delta P_{34}$, where the latter corresponds approximately to $P_3 - P_4$ (Fig.~\ref{fig1}). In Fig.~\ref{fig5}, we show relations between these quantities for sets of simulations in which $Q_1$ is fixed while $Q_2$ is varied.
We observe nonlinear relations between $r$ and $Q_3$ (Fig.~\ref{fig5}A), between $Q_3$ and the pressure loss along the chicane channel (Fig.~\ref{fig5}B), and between $r$ and the fluidic resistance of the chicane channel (Fig.~\ref{fig5}C). These nonlinear relations suggest a coupling between the pressure-flow relation of the chicane channel and the vortex size. 
We also note that discontinuities arise in the pressure-flow relation for the chicane channel (Fig.~\ref{fig5}B) that result from abrupt changes in the vortex size as $Q_2$ is varied (Fig.~\ref{fig5}A).
These discontinuities show the emergence of regions where the pressure-flow relation is negatively sloped, 
which  correspond to regions of negative \textit{differential} resistance.

\begin{figure*}[htb]
\centering 
\includegraphics[width=0.99\textwidth]{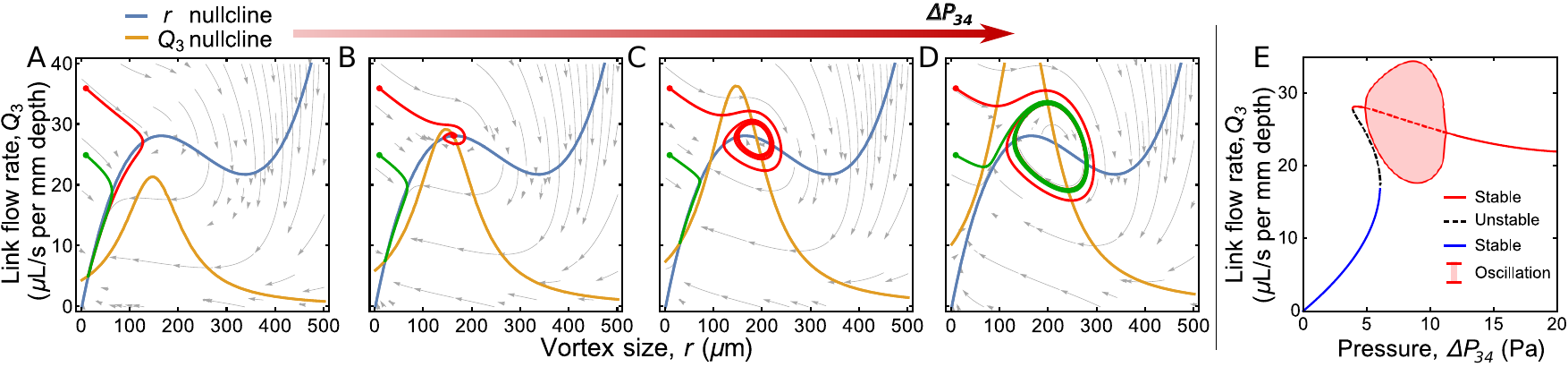}
\caption{
\textbf{Analytical dynamical model of flow through the chicane channel.}
(\textbf{A-D}) Phase space plots showing example trajectories (red and green curves) and streamlines (grey curves), generated from Eqs.~\ref{link1}-\ref{link2} for the flow rate and vortex dynamics at different values of $\Delta P_{34}$. Fixed point solutions to the equations exist at the intersections of the $r$ and $Q_3$ nullclines (i.e., the curves in phase space for which $\dot{r} = \dot{Q}_3=0$).   
The solution set may consist of one steady solution (A), three steady solutions (two stable, one unstable) (B), two steady solutions (one stable, one unstable) and a stable limit cycle (C), or a single stable limit cycle (D), depending on the value of $\Delta P_{34}$. (\textbf E) Bifurcation diagram of $Q_3$ produced for Eqs.~\ref{link1}-\ref{link2}. 
The parameters used here are: $Q_3^*=25\,\mu$L/s per mm depth, $r_b=146\,\mu$m, $r^*=250\,\mu$m, $L_b = 0.146$ cm, $\gamma=0.264\,\mu$m$^{-1}$, $\varepsilon=4166$ m$^{-1}$.
\vspace{-1mm}
}
\label{fig6}
\end{figure*}

\vspace{-2mm}
\section*{Analytical dynamical model}
\vspace{-2mm}
We now construct an analytical model of the system in Fig.~\ref{fig1} that characterizes our simulation results.
For unidirectional laminar flow through a straight channel, 
 the average flow rate of an incompressible fluid can be approximated from
 the Navier-Stokes equations as
\begin{equation}
l \dot{Q} = \Delta P - R Q,
\label{dyn1}
\end{equation}
where the dot implies a time derivative and $l$ may be referred to as the fluidic inductance \cite{Leslie2009}.
For   flow through a 
two-dimensional channel of length $L$,  where the characteristic time scale of the flow is larger than the viscous time scale, 
 the fluidic resistance and inductance  can be approximated as  $R=12 \mu L/w^3$ and $l=\rho L/w$.  
More generally, when the time scale of the flow exceeds the viscous time, memory effects in $R$ and $l$ become significant.
Under steady flow conditions, Eq.~\ref{dyn1} 
 reduces to $\Delta P = R Q$. 

One of the assumptions in the derivation of Eq.~\ref{dyn1} is that all streamlines of the channel flow are straight, which causes the nonlinear inertial terms in the Navier-Stokes equations to vanish.
Streamlines in the chicane channel clearly violate this assumption (Fig.~\ref{fig4}A) and nonlinear effects are therefore expected to be present. 
Indeed, we observe 
an approximately quadratic relation between the chicane channel resistance, $R_3$, and the vortex size, $r$, 
for $60< r <400\,\mu$m (Fig.~\ref{fig5}C).
To construct an approximate dynamical equation for $Q_3$, we use the form of Eq.~\ref{dyn1} with the constant resistance  replaced by a function of $r$. 
Specifically, we take $R_3(r) = 12 \mu (L_b +\gamma(r-r_b)^2)/w^3$, where $L_b$ serves as a base component of the resistance, $\gamma$ is a constant coefficient of the variable component that depends on the vortex size, and $r_b$ is the vortex size that minimizes the resistance (from Fig.~\ref{fig5}C, $r_b\approx 150\,\mu$m).
With this added dependence on $r$, we must also account for the dynamics of the vortex size. The steady-state relation between $Q_3$ and $r$ found through flow-controlled simulations (Fig.~\ref{fig5}A) can be well fit by a cubic equation of the form $Q_3 - Q_3^* = \eta (r-r^*)^3 - \xi (r-r^*)$, where $\eta$ and $\xi$ are positive parameters and $Q_3^*$ and $r^*$ are constants that shift the cubic relation from the origin. 
For simplicity, we consider the growth rate of $r$ to be proportional to the deviation from this equilibrium relation. Therefore, the dynamical equations that characterize the chicane channel  take the form
\begin{eqnarray}
\dot{r} &=& \varepsilon\big(Q_3 - Q^*_3 -  \eta(r - r^*)^3 + \xi(r-r^*)\big), \label{link1}\\
\dot{Q}_3  &=& \frac{w\Delta P_{34}}{\rho L_3}- \frac{12 \nu}{w^2 L_3}\big(L_b + \gamma(r-r_b)^2\big) Q_3, \label{link2}
\end{eqnarray}
where 
$\varepsilon$ is a positive constant. For suitable parameters, we find these equations capture the most salient properties in Fig.~\ref{fig4}, B and C. We show in Fig.~\ref{fig6} that for different $\Delta P_{34}$, Eqs.~\ref{link1}-\ref{link2} can exhibit bistability and stable limit cycle solutions.
We note that the additional dependence of the relations presented in Fig.~\ref{fig5} on $Q_1$ can be accounted for by allowing $\eta$ and $L_b$ to be functions of $Q_1$ (supplementary materials). 

A second nonlinear element of the network in Fig.~\ref{fig1} is the obstacle-laden channel. As the flow rate through this channel segment increases, stationary eddies form in the wake of the obstacles for moderate Reynolds numbers. The presence of many of these obstacles in close proximity 
generates large velocity gradients in the surrounding flow, which amplifies energy dissipation and results in 
an overall nonlinear pressure-flow relation for steady flow through the channel.
This equilibrium relation is well characterized by the Forchheimer equation used to describe steady flow through porous media, where inertial effects become significant  when $Re$ is of order 10  \cite{Andrade1999}. 
The Forchheimer equation takes the form $\Delta P =\alpha\mu L V + \beta\rho L V^2$, where $V$ is the average velocity, $\alpha$ is the reciprocal permeability, and $\beta$ is the non-Darcy flow coefficient. The latter two parameters are solely dependent on the system geometry, and not on the working fluid. For our two-dimensional channel with obstacles, we take $V= Q_4/w$ so that the pressure-flow relation for the channel becomes $P_4 =\alpha\mu L_4 Q_4/w + \beta\rho L_4 Q_4^2/w^2$, where $\alpha$ and $\beta$ are fit from simulations (supplementary materials and fig.~S4). 
We account for this nonlinearity in a dynamical equation for $Q_4$ by using a flow-rate-dependent function in place of the constant resistance in Eq.~\ref{dyn1}.
Specifically, we take $R_4(Q_4) = \alpha \mu L_4/w + \beta \rho L_4 Q_4/w^2$ so as to recover the Forchheimer equation in steady flow. 
A consequence of the nonlinearity of this channel is that it gives rise to a
non-monotonic relation between the pressure difference across the chicane channel and $P^{\mathrm{in}}$. As $P^{\mathrm{in}}$ is increased from zero, $Q_3$ initially increases, before decreasing, as indicated by the low-flow solution branch in Fig.~\ref{fig4}B. 

We now construct the dynamical model for the full network in Fig.~\ref{fig1} as follows: 
(i) we use flow relations of the form in Eq.~\ref{dyn1} with constant resistances 
for the three channel segments without obstacles and with a flow rate dependent resistance function (discussed above) for the obstacle-laden channel;  
(ii) we use Eqs.~\ref{link1}-\ref{link2} to describe the flow rate and vortex dynamics in the chicane channel with $\Delta P_{34}$ substituted by $(\zeta P_3-P_4)$, where $\zeta$ is a free parameter that may deviate from $1$ to account for an effective pressure difference across the chicane channel  due to the finite size of the channel junctions; and
(iii) we account for the most dominant minor pressure losses due to diverging flows at the channel junctions \cite{Crane1978}. For the latter, we include terms of the form $k Q_3 Q_5/Q_1$ in the flow equations for $Q_3$ and $Q_5$, where $k$ is a positive constant. This leads to six ordinary differential equations (five for flow rates and one for the vortex size), which can be reduced to four equations by making use of the equations that account for flow rate conservation at the channel junctions: $Q_1 = Q_3 + Q_5$ and $Q_2 = Q_4 - Q_3$
(see supplementary materials for details of the model).

The model predictions of the total flow rate, chicane channel flow rate, and vortex size for the network in Fig.~\ref{fig1} under a common driving pressure at the inlets are presented in supplementary materials fig.~\ref{fullPrediction}. The model captures well the complex solution structure observed in Figs.~\ref{fig2}A and ~\ref{fig4}, B and C, shows strong quantitative agreement with simulations, and provides several interpretations for the observed flow behavior.
First, spontaneous oscillations are found to arise through the transition from a fixed-point solution to a stable limit cycle via a supercritical Hopf bifurcation. The amplitude of the limit cycle grows with the driving pressure and eventually collides with the unstable solution surface of $Q_3$ and $r$, as shown in Fig.~\ref{fullPrediction}, thereby destabilizing the oscillating solution through a homoclinic bifurcation.
Second, the nonlinearity arising from the Forchheimer effect gives rise to the two distinct bistable regions (and thus two negative-conductance transitions), as a result of the non-monotonic relation between $P^{\mathrm{in}}$ and the pressure loss along the chicane channel. 
Third, the difference in the total flow rate between the solution branches is primarily determined by the minor losses. Without these terms, the model may still predict bistability, but 
the difference in total flow rate for solutions belonging to different branches would be negligible.

Our model can also be used to integrate the nonlinear behaviors described above into larger microfluidic systems. As an example, we consider an extended network with three outlets (supplementary material,  fig.~\ref{mixing}), with two separate inlet flows.
By driving the flows through this network using a common pressure, three unique oscillatory flow compositions can be realized at the outlets. Our model predictions show that the flow composition at the individual outlets is different, but the flow rate at the outlets oscillate in phase (supplementary materials, fig.~\ref{mixing}B). Thus, the property that flow rates through all channel segments oscillate with the same period can be extended to larger networks and used to produce synchronized, time-dependent output flow patterns.

\vspace{-2mm}
\section*{Discussion}
\vspace{-2mm}
 Motivated by the challenge of developing built-in controls in microfluidics, we identified mechanisms that can facilitate integration without dependence on movable parts or external actuation (other than through the working flow). This includes our demonstration of self-sustained oscillations, which can be used for timing and synchronization of  flows through different channels;  multistability and associated transitions, which can be used for signal amplification and switching, and hysteresis; which could serve as a possible mechanism for memory. In particular, we demonstrated the emergence of spontaneous periodic variations in the relative uptake rates from different inlets, which can be explored to generate time-dependent mixtures and output flow patterns.  While these dynamical behaviors  may resemble those found in microelectronics, they rely on effects that do not have direct analogs in electrical networks, namely fluid inertia and the resulting nonlinearity arising from interactions between components.

 Our results demonstrate that
fluid inertia effects can be amplified
and induce behaviors in fixed-structure microfluidic systems that have not been previously generated without external actuation. Indeed, the negative-conductance transitions, spontaneous oscillations, hysteresis, and multistability simulated and modeled in our system all emerge as a consequence of coupling between the geometric structure of the network and fluid inertia effects.
Flows around obstacles and through the porous-like channel are determinant for generating these dynamics. Porous media microfluidics have become important for the study of flows through natural systems  and lab-controlled experiments \cite{Anbari2018}.  In this work, we placed new emphasis on the viability of porous-like structures to serve as nonlinear fluid resistors and harness fluid inertia effects for non-local flow control throughout the network, which, crucially, can be realized as a built-in mechanism.
 Given that our system can be constructed from rigid materials, it is able to withstand a wide range of driving pressures  (e.g., $1\,$Pa--$10^6\,$ Pa), which facilitates implementation across the length scales relevant to microfluidics.

The flow dynamics that arise in our system can be tailored for various applications.
Microfluidic systems capable of
carrying out sequential operations generally require a timing mechanism that is either generated from an external device or through the use of flexible valves \cite{Duncan2013}. The oscillations that arise in our system could serve as an on-chip frequency reference and enable process synchronization or waveform synthesis. Moreover, the vortex dynamics that give rise to the oscillations may be used to enhance state-of-the-art methods for particle sorting and manipulation that function through interactions between particles and micro-vortices \cite{Dhar2018}. In particular, vortex dynamics can be used to produce complex (and even chaotic) particle trajectories in laminar flows \cite{Karolyi2000}. 
Finally, microfluidic networks are now widely used in the study of colloids \cite{Desmond2015} and active matter \cite{Woodhouse2017}. Our system offers a rich environment to further investigate these materials given that they exhibit surprising collective behavior when placed in different flow fields \cite{Han2019} and when driven through porous media \cite{Shin2017}. Moving forward, 
we anticipate
that the coupling of fluid inertia effects and network geometry can be further explored across microfluidic applications to yield new built-in flow control functionality.

\section*{Materials and Methods}
\subsubsection*{Navier-Stokes simulations}
Our simulations of the Navier-Stokes equations for incompressible fluid were performed using OpenFOAM-version 4.1. Meshes of the system geometry were generated using Gmsh-version 2.9.3, with average cell area ranging from $10$ to $70\,\mu$m$^2$. The pisoFoam and simpleFoam solvers in OpenFOAM were used for time-dependent and steady-state simulations, respectively.
For simulations where a Dirichlet static pressure boundary condition was used at an inlet/outlet, a Neumann boundary condition was used to set the gradient of the velocity field to zero in the direction normal to the inlet/outlet. This combination of boundary conditions results in a fully-developed velocity profile at the inlet/outlet, and corresponds to the physical situation in which the channels extend upstream and downstream of the computational domain.
Similarly, 
when instead the flow rate was controlled at an inlet, a parabolic velocity profile was specified and a zero-gradient boundary condition was used for the pressure. 
Values for $P_3$ and $P_4$ in Fig.~\ref{fig5} were measured by averaging the pressure sampled across the channel width at a distance $3w/5$ downstream of the chicane channel junctions.

\subsubsection*{Network dimensions}
For the network presented in Fig.~\ref{fig1}, the individual channel segment lengths, as labeled in Fig.~\ref{fig1}D, are:
$L_1 = 0.1$, $L_2 = 0.6$, $L_3 = 0.1$, $L_4 = 1.0$, and $L_5 = 0.5$, all in cm. The cylindrical obstacles in the obstacle-laden channel (Fig.~\ref{fig1}C) are separated by a distance of approximately $6w/5$, where $w = 500\,\mu$m is the channel width (common to all channels in the network). The blade-like barriers in the chicane channel (Fig.~\ref{fig1}B) are each placed a distance $w/2$ from the midpoint of the axis along the channel.

\subsubsection*{Reynolds numbers} The characteristic length scale used in defining the Reynolds numbers of the flows is the hydraulic diameter of the channels, defined as $4 A/P$, where $A$ is the area and $P$ is the perimeter of the channel cross section (common to all channel segments). In two dimensions, the hydraulic diameter is $2w$ and the characteristic velocity used is $Q/w$. Therefore, we define the Reynolds number for individual channel segments to be $2\rho Q/\mu$, where $Q$ is the associated flow rate through the channel.

\bibliographystyle{Science}




\section*{Acknowledgments}
\textbf{Funding:} This work was supported by ARO Grant No.\ W911NF-15-1-0272 and a Northwestern University Presidential Fellowship.\\ 
\textbf{Competing interests:} The authors declare no competing interests.\\
\textbf{Data and material availability:} All supporting data and materials are available upon request, by contacting A.E.M.\\

\section*{Supplementary materials}
Supplementary Text\\
Figs. S1 to S7\\
Movie S1\\

\newpage
\clearpage

\setcounter{page}{1}
\renewcommand{\theequation}{S\arabic{equation}}
\setcounter{equation}{0}
\renewcommand{\thefigure}{S\arabic{figure}}
\setcounter{figure}{0}
\renewcommand{\thetable}{S\arabic{table}}
\setcounter{table}{0}
\renewcommand{\vec}{\mathbf}

\renewcommand{\thesection}{S\arabic{section}}

\begin{center}
{\bf\sc\LARGE Supplementary Material}

{\it Spontaneous oscillations and negative-conductance transitions in microfluidic networks
}

Daniel J. Case, Jean-R\'egis Angilella, and Adilson E. Motter
\end{center}

\tableofcontents

\bigskip
\bigskip
\bigskip
\section{Supplementary Text}
\subsection{Supporting results on direct fluid dynamics simulations}

\subsubsection*{Flow-controlled Navier-Stokes simulations}\label{S32}
\begin{figure}[h!!] 
\centering
\includegraphics[width=0.6\textwidth]{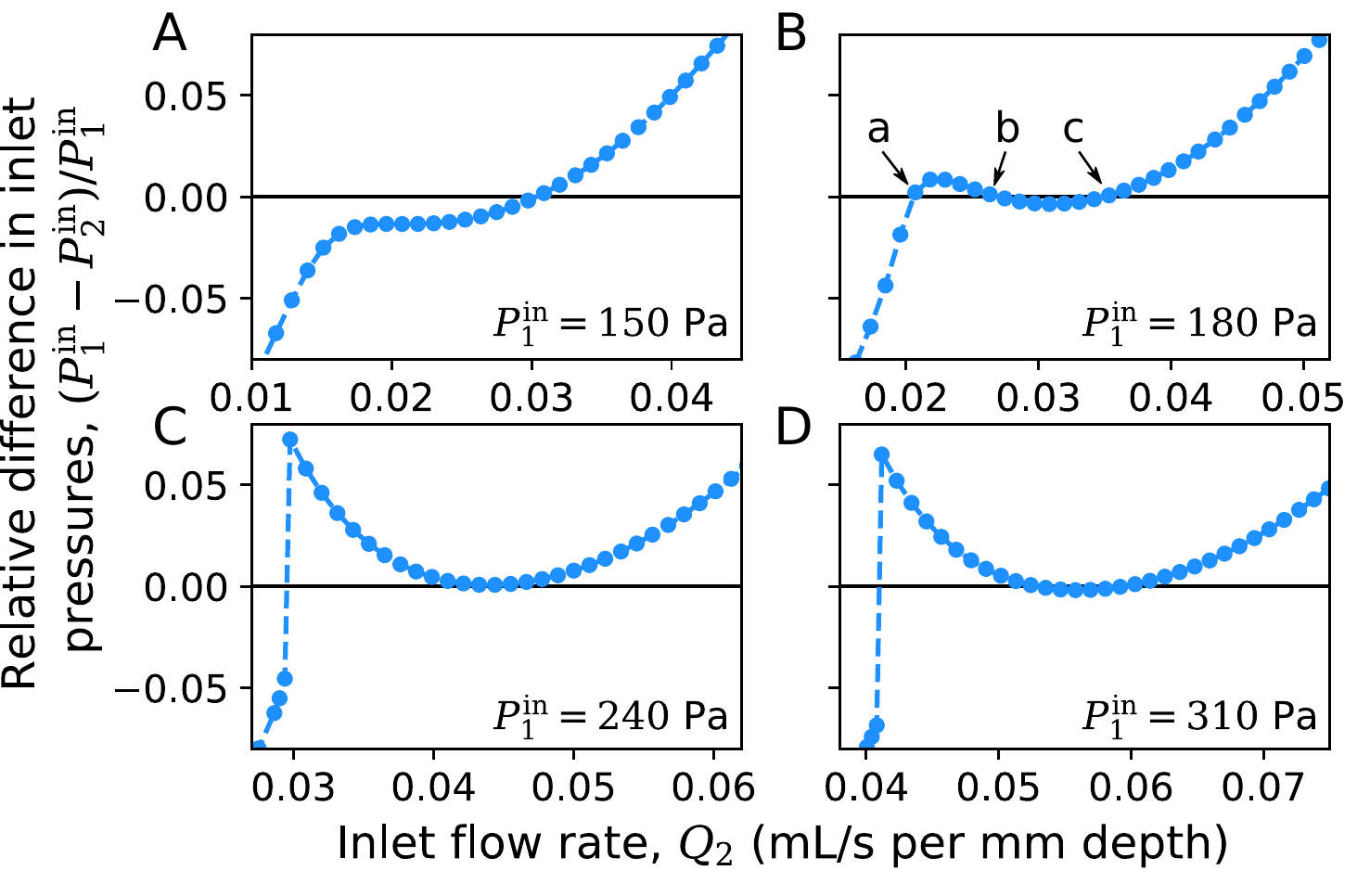}
\caption{
\textbf{Flow-controlled Navier-Stokes simulations reveal bistability.}
(\textbf{A - D}) Relative difference in inlet pressures of the network in Fig.~1 for fixed $P^{\mathrm{in}}_1$ with $Q_2$ varied.
In accordance with the two bistable regions identified in Fig.~2A, the number of axis crossings, as $P^{\mathrm{in}}_1$ is fixed to higher values, varies between one (A and C) and three (B and D).
The total flow rates corresponding to the solutions labeled in (B) are marked in Fig.~2A.
}
\label{PressureSampSI}
\end{figure}

To show how multistability arises for pressure-driven flows through the network in Fig.~1, we present additional simulations in which  pressure is controlled at one inlet ($P^{\mathrm{in}}_1$), flow rate is controlled at the other inlet ($Q_2$), and the resulting value of $P^{\mathrm{in}}_2$ is measured. In fig.~\ref{PressureSampSI}, we show the relative difference between the inlet pressures for several examples in which $P^{\mathrm{in}}_1$ is fixed with $Q_2$ being varied. Points where this difference vanishes (i.e., crossings of the horizontal axis) indicate potential solutions for the case in which $P^{\mathrm{in}}_1$ and $P^{\mathrm{in}}_2$ are controlled to be equal. This is an indirect way of identifying both the stable and unstable solutions for the pressure-controlled simulations since, by their very nature, the latter are not directly observable. Indeed, we see for low driving pressure (fig.~\ref{PressureSampSI}A) a single solution exists, and as the pressure is increased, two additional solutions emerge through a saddle-node bifurcation. The three solutions indicated by a, b, and c in fig.~\ref{PressureSampSI}B correspond to the labeled solutions in Bistable Region I in Fig.~2A. For a higher driving pressure (fig.~\ref{PressureSampSI}C), the only indicated solution for which pressures at the inlets are equal occurs over a discontinuity and corresponds to a driving pressure in Fig.~2A that yields a single unsteady solution. For still higher pressures (fig.~\ref{PressureSampSI}D), multiple solutions reemerge and correspond to solutions within Bistable Region II in Fig.~2A.
We determine the stability of all identified solutions by performing time-dependent simulations.

\subsubsection*{Flow-rate and vortex oscillations}
\begin{figure}[h!!] 
\centering
\includegraphics[width=0.6\columnwidth]{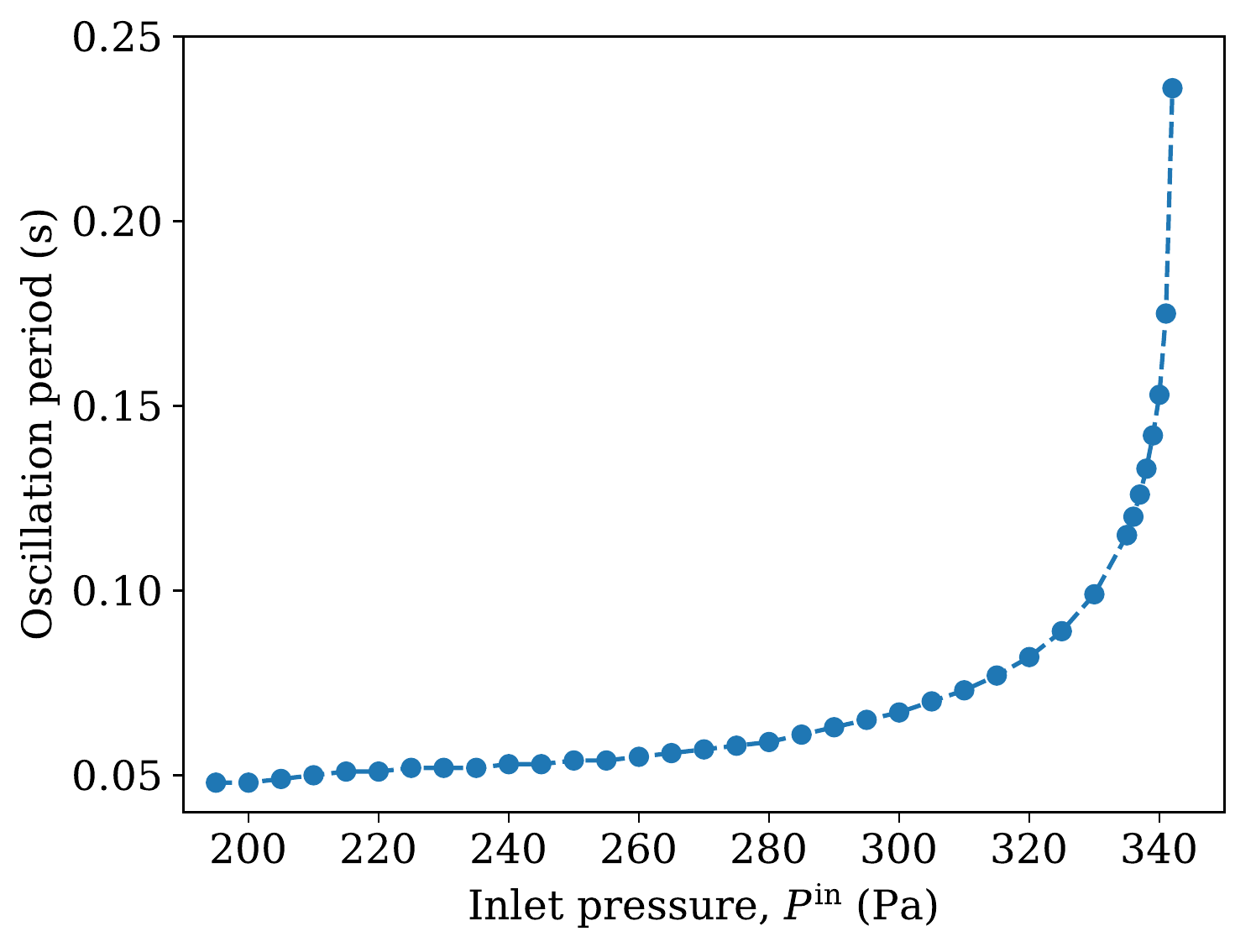}
\caption{\textbf{Divergence of oscillation period.} Period of oscillation of unsteady solutions along the high-flow branch in Fig.~2A
for different values of $P^{\mathrm{in}}$.
 The dots represent the results from Navier-Stokes  simulations, and the interpolating dashed line is a guide to the eye.
}
\label{periods}
\end{figure}

For unsteady solutions along the high-flow branch in Figs.~2A and 4, B and C, all variables (flow rates and vortex size) oscillate with the same period. The period of the oscillating solutions as a function of the driving pressure $P^{\mathrm{in}}$ is presented in fig.~\ref{periods}. As the driving pressure approaches the value at which the unsteady solutions become unstable, the period of the oscillations diverges, which is indicative of a homoclinic bifurcation, whereby the stable limit cycle collides with the unstable solution surface.

\begin{figure}[h!!] 
\centering
\includegraphics[width=0.55\columnwidth]{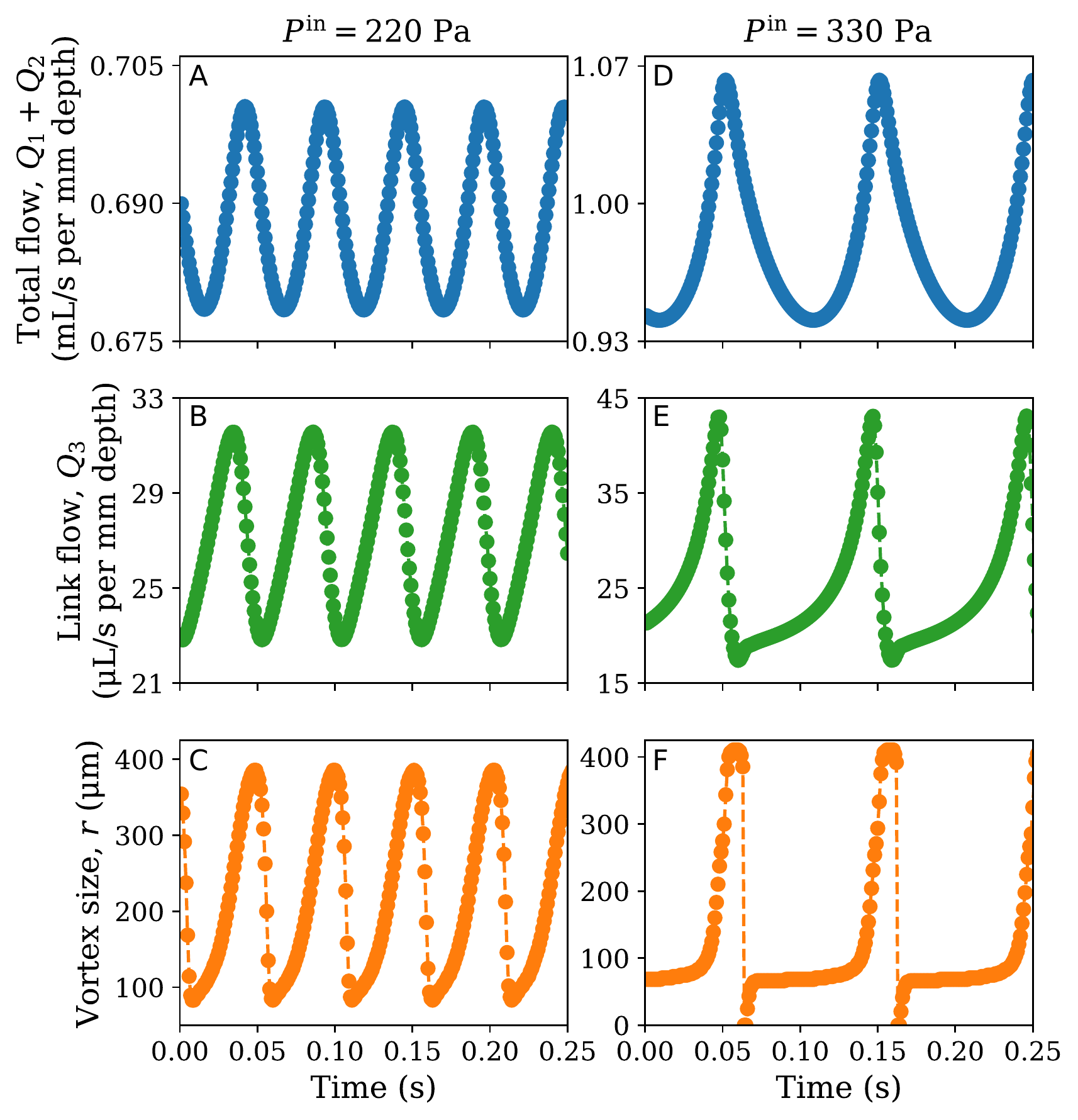}
\caption{\textbf{Example time series of oscillating solutions determined by Navier-Stokes simulations.} (\textbf A and \textbf D) Oscillations in the total flow rate, (\textbf B and \textbf E) flow rate through the chicane channel, and (\textbf C and \textbf F) vortex size for two different fixed driving pressures, $P^{\mathrm{in}} = 220\,$Pa (A-C) and  $P^{\mathrm{in}} = 330\,$Pa (D-F).
The symbols represent measurements from direct fluid dynamics simulations, sampled uniformly in time, and the interpolating line is used to guide the eye.
}
\label{oscSeries}
\end{figure}

Example time series for the total flow rate, vortex size, and flow rate through the chicane channel are presented in fig.~\ref{oscSeries} for two values of $P^{\mathrm{in}}$. Two timescales are particularly evident in the oscillations of the flow rate through the chicane channel (fig.~\ref{oscSeries}, B and E) and the vortex size (fig.~\ref{oscSeries}, C and F), whereby a comparatively slow growth in each of these quantities is followed by a rapid decline.
As the amplitude of the oscillations in the flow rate grow along the high-flow solution branch, so does the amplitude in the oscillations of the vortex size. For high enough driving pressures, the vortex vanishes for a very short time relative to its period, as shown in the time series of $r$ in fig.~\ref{oscSeries}F.
To better facilitate the visualization of the oscillation amplitude in Fig.~4C, 
the maximum amplitude value was taken to be the maximum value of $r$ over one oscillation period at each driving pressure, while the minimum was taken to be the value of $r$ at which the magnitude of $dr/dt$ was minimized (other than at the extrema of $r$). This adjusted minimum represents the minimum vortex size outside of the the very short time range in which the vortex vanishes.

\subsection{Dynamical model of microfluidic network}\label{S1}
~\\
The model describing flows through the network in Fig.~1 is based on Eqs.~1-3 in the main text. For the flow equations of $Q_1$, $Q_2$, and $Q_5$, we use the form in Eq.~1, where the resistances take the standard form for a two-dimensional channel, $12\mu L_i/w^3$. The corresponding length of each channel segment is labeled in Fig.~1D.

\subsubsection*{Chicane channel flow and vortex dynamics}\label{S11}
To characterize the flow rate through the chicane channel, we use Eqs.~2-3, which couple the pressure-flow relation and vortex dynamics. We account for the additional dependence of Eq.~2 on $Q_1$, as observed in Fig.~5A, by allowing $\eta$ to be a function of $Q_1$. To keep the model simple, we define 
\begin{equation}\label{etaEq}
\eta(Q_1)=p+q\,Q_1, 
\end{equation}
where $p$ and $q$ are constants. Further, we set 
\begin{equation}\label{xiEq}
\xi = \frac{(r^*)^3\eta - Q^*}{r^*},
\end{equation}
which ensures that $r=0$ for $Q_3=0$, corresponding to the physical requirement that the vortex vanishes as the flow rate goes to zero. 

The pressure drop $\Delta P_{34}$ in Eq.~3 corresponds to the pressure loss along the chicane channel. This loss is approximately equal to $P_3 - P_4$ (Fig.~1A). However, the pressure field can vary significantly over short distances at the channel junctions. To account for this effect, as well as the fact that the chicane channel has a comparatively wide width and short length, we  define the effective pressure drop as $\zeta P_3 - P_4$, where $\zeta$ is a constant expected to be near one. We also note that the measured resistance of the chicane channel (Fig.~5C) also shows dependence on $Q_1$. This can be included in the model by allowing $L_b$ in Eq.~3 to vary with $Q_1$, but this modification did not present marked differences in the model prediction, so we chose not to include it here.

\subsubsection*{Flow through obstacle-laden channel}\label{S12}
\begin{figure}[htb] 
\centering
\includegraphics[width=0.6\columnwidth]{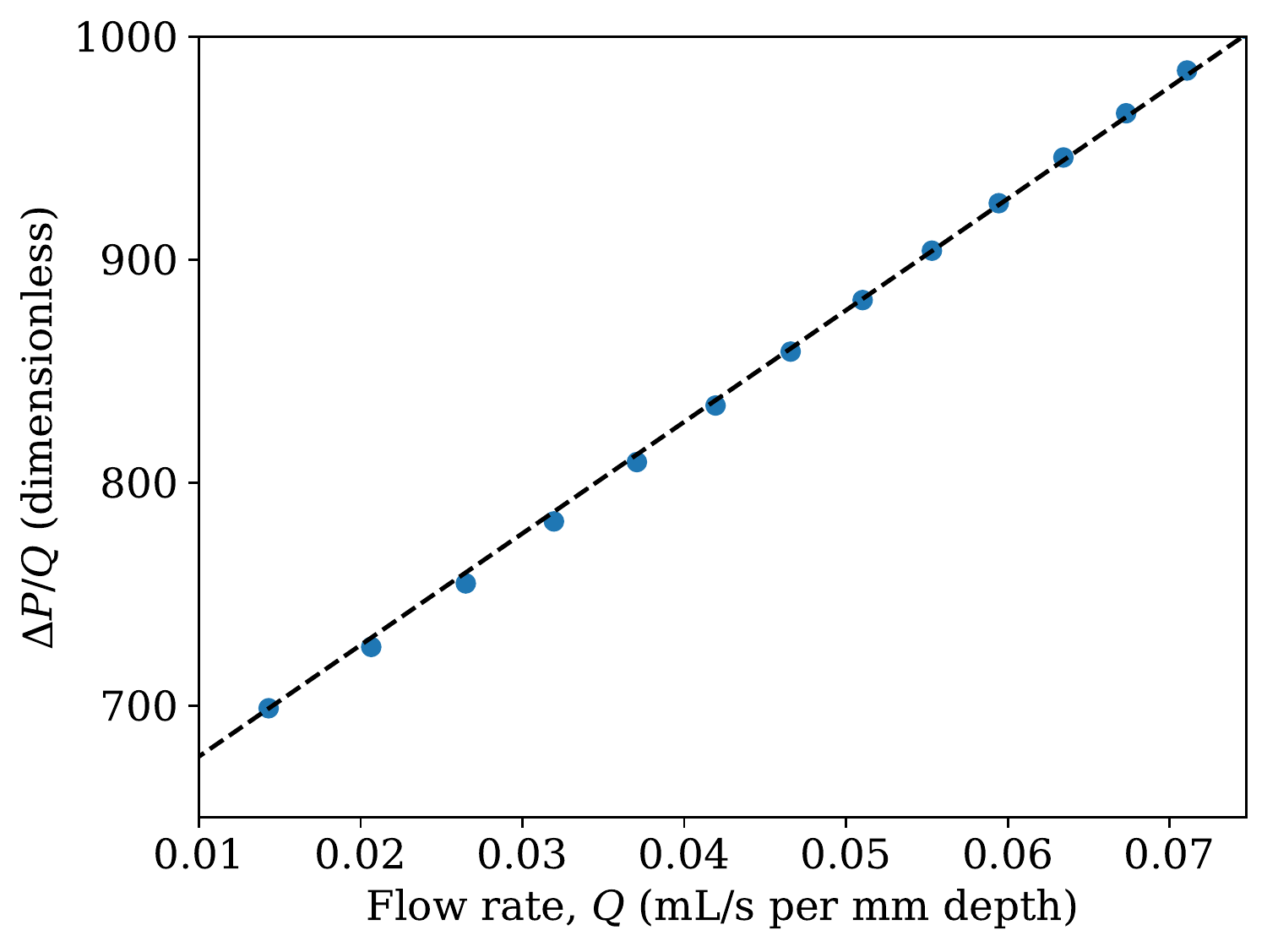}
\caption{\textbf{Nonlinear flow through channel with obstacles. }
The Navier-Stokes simulation results (symbols) show the relation between the fluidic resistance, $\Delta P/Q$, and the flow rate, $Q$, for a channel with six cylindrical obstacles. The linear fit (dashed line) confirms that $\Delta P$ is well approximated by Eq.~\ref{Forch}. The channel length is $1\,$cm, and the fitted values of $\alpha$ and $\beta$ are $1.24\times10^{8}\,$m$^{-2}$ and $513\,$m$^{-1}$, respectively. The resistance is non-dimensionalized by dividing it by $\mu/w^2$.
}
\label{CylChannel}
\end{figure}

The channel with flow rate $Q_4$ contains six cylindrical obstacles. For steady flow, the pressure-flow relation is well-characterized by the Forchheimer equation, as discussed in the main text, whereby the pressure loss along the channel depends quadratically on the flow rate. The steady pressure-flow relation takes the form
\begin{equation}\label{Forch}
\Delta P = \frac{\alpha \mu L}{w} Q + \frac{\beta \rho L}{w^2}Q^2.
\end{equation}
In fig.~\ref{CylChannel}, we show that our simulation results conform to this equation for the isolated obstacle-laden channel. The observed linear relation between $Q$ and $\Delta P/Q$ with non-zero slope indicates an approximately quadratic relation between $Q$ and $\Delta P$. This nonlinearity arises as a result of
large velocity gradients and eddies that
form in the wakes of the obstacles as the flow rate is increased, which thereby alters the fluidic resistance of the channel. 
Therefore, we approximate the dynamic flow-rate relation for $Q_4$ by using Eq.~1 with the resistance $R$ replaced by the flow-rate-dependent expression $R_4(Q_4) = \alpha \mu L_4/w + \beta \rho L_4 Q_4/w^2$.

\newpage
\subsubsection*{Minor losses at channel junctions}\label{S13}
The Reynolds numbers of the flows through the system are up to  
 the order of $100$  for the chicane and obstacle-laden channels and up to the order of $1000$ for obstacle-free channels. Given that nonlinearity arises in the pressure-flow relations for different channels, fluid inertia effects are clearly significant. Therefore, additional pressure losses in areas where streamlines combine, diverge, or bend---so called minor losses---should be considered.
The flow rates $Q_1$ and $Q_5$ are an order of magnitude higher than elsewhere in the network, since the absence of obstacles in the associated channels yield low fluidic resistances.
 As a result, the most dominant minor losses are expected to occur at the junction of $Q_1$, $Q_3$, and $Q_5$.
Minor loss terms are typically found empirically for different geometries, and at flow junctions they generally take the form of a scaling factor that depends on the flow rate with a coefficient that is a function of the ratio of diverging flows (36). The scaling factor and coefficient can each be nonlinear, in principle, but for our model we use the simple expression $k Q_3 Q_5/Q_1$, where $k$ is a constant. A term of this form is included in the flow rate equation for $Q_3$ and $Q_5$, with independent values of $k$.

\subsubsection*{Non-dimensional model equations}\label{S14}
Our model of the microfluidic network in Fig.~1 is constructed from equations describing each component of the network and includes terms that account for interactions between components, as described in the preceding sections.
The variables and parameters in the model can be non-dimensionalized using the following definitions:
\begin{equation}
\begin{split}
\overline{P}&=\frac{P}{\rho \nu^2/w^2} ;\, \overline{Q}=\frac{Q}{\nu} ;\, \overline{L}=12\frac{L}{w};\, \overline{r}=12\frac{r}{w};\, \overline{\gamma}=\frac{\gamma w}{12} ;\,  \overline{t} = t\frac{\nu}{w^2} ;\\ \overline{\eta} &= \frac{\eta}{\nu}\bigg(\frac{w}{12}\bigg)^3;\,
\overline{p} = \frac{p}{\nu}\bigg(\frac{w}{12}\bigg)^3 ;\, \overline{q} = q\bigg(\frac{w}{12}\bigg)^3 ;\,
 \overline{\xi}=\xi\frac{w}{12 \nu};\, \overline{\alpha}=w^2 \alpha, \overline{\beta}=w \beta;\, \overline{\varepsilon}=12 w \varepsilon;
\end{split}
\end{equation}
where $t$ indicates time and $\nu = \mu/\rho$ is the kinematic viscosity.

The complete set of non-dimensional equations that define the model read:
 \begin{eqnarray}
\dot{Q}_1 &=& 12\Bigg( \frac{P_1^{\mathrm{in}} - P_3}{L_1} - Q_1\Bigg), \label{model1}\\
\dot{Q}_2 &=& 12\Bigg( \frac{P^{\mathrm{in}}_2 - P_4}{L_2} - Q_2\Bigg),\label{model2}\\
\dot{Q}_3 &=& 12\Bigg( \frac{\zeta P_3 - P_4}{L_3} - \frac{1}{L_3}\big(L_{\mathrm{b}} + \gamma (r - r_b)^2\big) Q_3 + k_3 \Big(\frac{Q_5}{Q_1}\Big) Q_3 \Bigg), \label{model3}\\
\dot{Q}_4 &=& 12\Bigg( \frac{P_4}{L_4} - \frac{1}{12}(\alpha + \beta Q_4) Q_4\Bigg) \label{model4},\\
 \dot{Q}_5 &=& 12\Bigg( \frac{P_3}{L_5} - Q_5 + k_5 \Big(\frac{Q_3}{Q_1}\Big) Q_5 \Bigg), \label{model5}\\
 \dot{r} &=& \varepsilon \big( (Q_3 - Q_3^*) - \eta(Q_1) (r-r^*)^3 + \xi (r-r^*)\big),\label{model6}\\
 Q_{1} &=& Q_3 + Q_5, \label{model7}\\
 Q_{2} &=& Q_4 - Q_3 \label{model8},
 \end{eqnarray}
where the bars over non-dimensional quantities are omitted for brevity.
  
\begin{figure}[tb] 
\centering
\includegraphics[width=0.6\columnwidth]{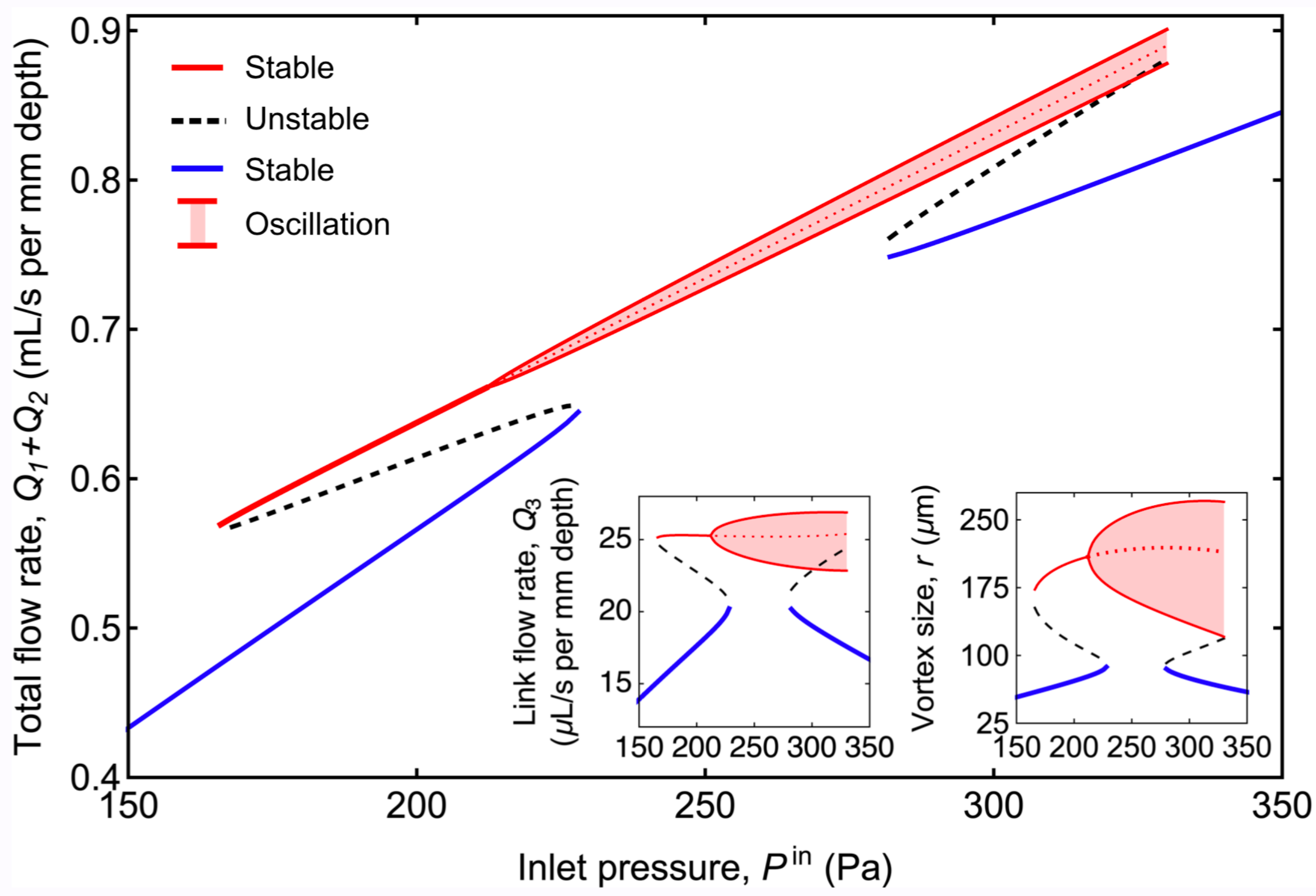}
\caption{
\textbf{Model predictions of network flow dynamics.} Bifurcation diagrams produced by our model for the total flow rate, flow rate through the chicane channel (left inset), and vortex size (right inset). Model parameters were fit to simulation results.
}
\label{fullPrediction}
\end{figure}

For $\eta(Q_1)$ and $\xi$ in Eq.~\ref{model6}, we use non-dimensionalized forms of Eqs.~\ref{etaEq}-\ref{xiEq}. In view of  Eqs.~\ref{model7}-\ref{model8}, the model can be reduced to only four equations for the variable $Q_1$, $Q_2$, $Q_3$, and $r$.  We show the model prediction of the system dynamics in fig.~\ref{fullPrediction}.

\subsubsection*{Model parameters}\label{S15}
The undeclared parameter values we use for the model predictions in fig.~\ref{fullPrediction} are as follows. 
For the channel widths and lengths, as labeled in Fig.~1D, we use the same values as in the simulations.
The fitted values of $\alpha$ and $\beta$ are declared in the caption of fig.~\ref{CylChannel}.
The remaining non-dimensional parameters are:
\begin{equation} \label{params1}
\begin{split}
\zeta &= 1.08, \; \overline{L}_b=42, \; \overline{\gamma}=7.35, \; \overline{r}_b = 4.1, \; \overline{r}^* = 5.5, \; \overline{\varepsilon}=10.9,\\  \overline{k}_3 &= 15.75, \;  \overline{k}_5 =12, \; \overline{Q}^{\,*}_3 = 25, \;
\overline{p} = 0.12, \; \overline{q}=0.8\times 10^{-5},
\end{split}
\end{equation}
where $\overline{p}$ and $\overline{q}$ are used in the non-dimensional form of Eq.~\ref{etaEq}.
 
\subsubsection*{Three-outlet network} 
\begin{figure}[tb] 
\centering
\includegraphics[width=0.75\columnwidth]{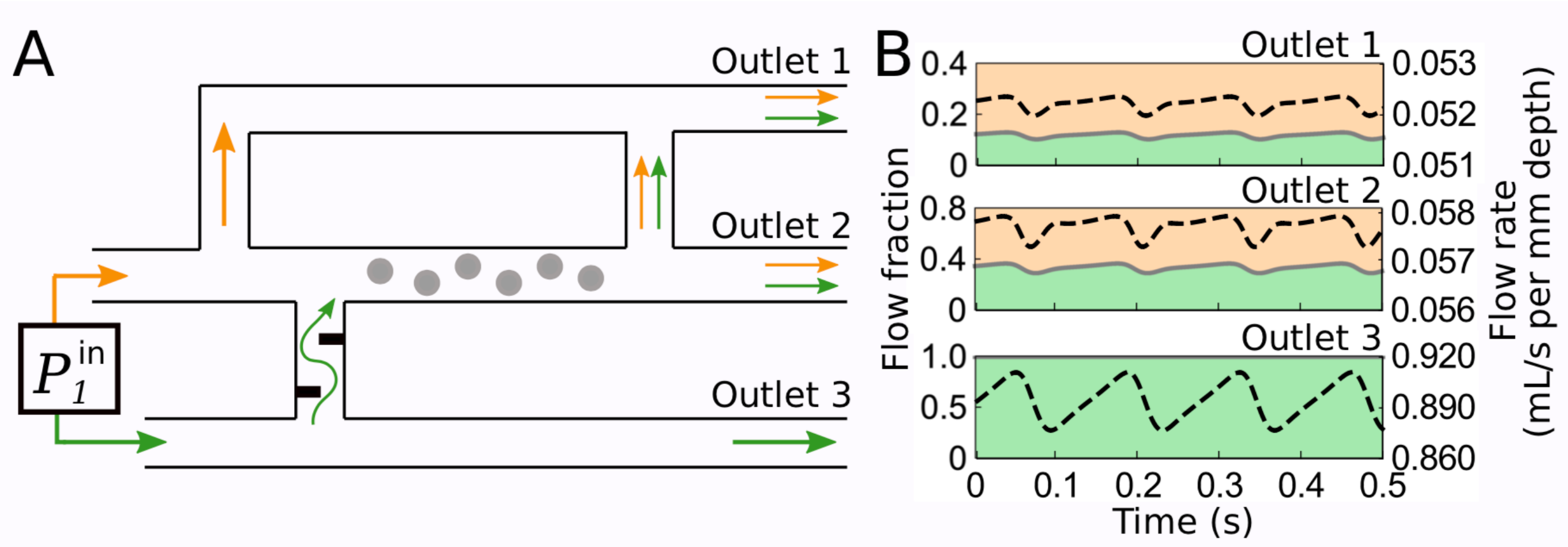}
\caption{
\textbf{Synchronous output patterns.}
(\textbf A) Schematic of extended microfluidic network with three outlets, where inlet flows are marked orange and green, the inlet pressure is held fixed at  $P_\mathrm{in} = 350\,$ Pa, and the outlet pressure is held fixed at zero. (\textbf B) Model predictions for the network shown in (A), where the green areas show the portions of the flows that originate from the bottom inlet (left axis) and the dashed lines indicate the total flow rate through the respective outlet (right axis).}
\label{mixing}
\end{figure}

We demonstrate how different output flow patterns can be generated in an expanded version of the network detailed above. Specifically, we consider the scenario in which three additional channel segments without obstacles are incorporated into the network described in Fig.~1, resulting in a system with three outlets, shown in fig.~\ref{mixing}.

\begin{figure}[htb] 
\centering
\includegraphics[width=0.7\columnwidth]{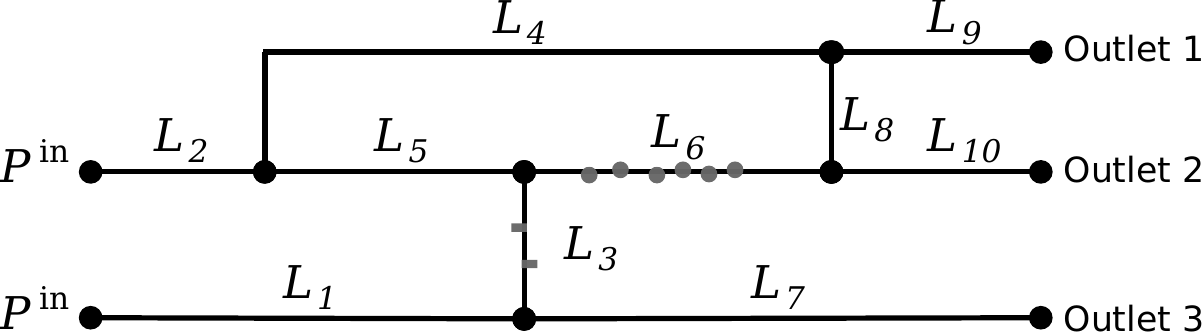}
\caption{\textbf{Network schematic of three-outlet system in fig.~\ref{mixing}A.} The length of each channel segment is denoted by $L_i$.
\vspace{1cm}
}
\label{threeOutNetwork}
\end{figure}

A schematic of the three-outlet network is presented in fig.~\ref{threeOutNetwork}. The predictions of the outlet flows presented in fig.~\ref{mixing}B are derived from a dynamic model constructed in the same manner as in Eqs.~\ref{model1}-\ref{model8}. The model consists of ten flow rate equations (one for each channel segment), five flow-rate conservation equations, and one equation for the vortex dynamics. Aside from the channel segment lengths, all parameter values used are the same as in Eq.~\ref{params1}. The channel segment lengths used for the prediction in fig.~\ref{mixing}B and corresponding to the labels in fig.~\ref{threeOutNetwork} are: $L_1 = 0.1035$, $L_2 = 0.3$, $L_3 = 0.1$, $L_4 = 10.0$, $L_5 = 0.3$, $L_6 = 1.0$, $L_7 = 0.45$, $L_8 = 0.1$, $L_9 = 0.02$, $L_{10} = 0.05$, all in cm. Finally, the driving pressure used was $P_\mathrm{in}=350\,$ Pa, which yielded oscillating flows throughout the network.

\newpage
\section{Supplementary Movie}

\bigskip
\noindent
\textbf{Movie S1.} Streamlines of the flow through the chicane channel illustrate oscillations in the flow rates and vortex size. The static pressure at the inlets is held constant at $220\,$ Pa. The arrows indicate the average flow direction and the unit of time is seconds.

\end{document}